\documentclass[12pt]{article}
\usepackage{epsfig,amsfonts,amssymb}
\begin{document}

\newcommand{\be}{\begin{equation}}
\newcommand{\ee}{\end{equation}}
\newcommand{\bea}{\begin{eqnarray}}
\newcommand{\eea}{\end{eqnarray}}
\newcommand{\beas}{\begin{eqnarray*}}
\newcommand{\eeas}{\end{eqnarray*}}

\parskip 12 pt

\begin{titlepage}
\begin{flushright}
{\small CU-TP-1052} \\
{\small KIAS-P02014} \\
{\small hep-ph/0204120}
\end{flushright}

\begin{center}

\vspace{2mm}

{\Large \bf QCD vacuum structure in strong magnetic fields}

\vspace{3mm}

Daniel Kabat${}^a$\footnote{electronic mail: kabat@phys.columbia.edu},
Kimyeong Lee${}^{ab}$\footnote{electronic mail: klee@kias.re.kr}
and Erick Weinberg${}^a$\footnote{electronic mail: ejw@phys.columbia.edu}

\vspace{1mm}

${}^a${\small \sl Department of Physics} \\
{\small \sl Columbia University, New York, NY 10027}

\vspace{1mm}

${}^b${\small \sl School of Physics, Korea Institute for Advanced Study} \\
{\small \sl Cheongryangri-Dong, Dongdaemun-Gu, Seoul 130-012, Korea}

\end{center}

\vskip 0.3 cm

\noindent
We study the response of the QCD vacuum to strong magnetic fields,
using a potential model for the quark-antiquark interaction.  We find
that production of spin-polarized $u\bar{u}$ pairs is energetically
favorable for fields $B > B_{\rm crit} \sim 10 \,\, {\rm GeV} {}^2$.
We contrast the resulting $u\bar{u}$ condensate with the quark
condensate which is present at zero magnetic field, and we estimate
the corresponding magnetization as a function of $B$.

\end{titlepage}

\section{Introduction}

Strong magnetic fields are interesting from several perspectives.
From a theoretical point of view, an external magnetic field allows
one to probe the vacuum structure and correlation functions of a
quantum field theory.  Strong magnetic fields are also of interest in
astrophysics.  There may be neutron stars with fields of up to
$10^{14}$ -- $10^{15}$ Gauss \cite{magnetar}, while it has been
suggested that much larger fields existed in the early universe
\cite{Primordial}.

With increasing magnetic field, the first place one might expect
something interesting to happen is at the scale set by the electron mass,
\begin{equation}
B = m_e^2 / \sqrt{\alpha} = 4.4 \times 10^{13} \,\, {\rm Gauss}\,.
\end{equation} 
At this scale an electron's Landau energy equals its rest energy.
Magnetic fields of this strength have a significant effect on atomic
and molecular physics, as reviewed in \cite{Duncan}.  However the
structure of the QED vacuum does not change dramatically in fields of
this magnitude.  Corrections to the energy of a free electron in the
lowest Landau level are small, proportional to the electron's
anomalous magnetic moment, so it is not energetically favorable to
produce $e^+e^-$ pairs.  The binding energy of positronium is small
and does not change this conclusion, so the QED vacuum is
stable.\footnote{This is discussed in more detail in \cite{Duncan}.
Exponentially large magnetic fields, in contrast, have been shown to
catalyze chiral symmetry breaking in QED \cite{DCSB}.}

The next place one might expect something interesting to happen is at
the QCD scale,
\begin{equation}
B = \Lambda_{QCD}^2/\sqrt{\alpha} \approx 10^{19} \,\, {\rm Gauss}\,.
\end{equation} 
This regime will be the main focus of this paper.  We will argue that
the perturbative QCD vacuum becomes unstable to formation of a
quark-antiquark condensate.  The basic physics is easy to understand:
the strong magnetic field restricts quarks and antiquarks to move in
one dimension, and the strongly attractive QCD potential then leads to
formation of a bound state with negative energy.  We will argue that
at sufficiently large magnetic fields the effective coupling becomes
weak and perturbative QCD can be used.

The next interesting regime starts at the electroweak scale,
\begin{equation}
B = m_W^2 / \sqrt{\alpha} \approx 10^{24} \,\, {\rm Gauss}\,.
\label{electroweakfield}
\end{equation} 
A field of this magnitude has been argued to drive electroweak
symmetry restoration \cite{Electroweak}.  Finally, a grand unified
theory with magnetic monopoles of mass $M_{\rm mon}$ could provide an
absolute upper bound on a possible magnetic field.  Extrapolation of
semiclassical calculations \cite{Affleck:1981ag} suggests that monopole
pair production would become copious and short out any existing
magnetic field when
\begin{equation}
      B \sim \sqrt{\alpha} M_{\rm mon}^2 \sim 10^{52}
 \left({M_{\rm mon}\over 10^{17} \,\, {\rm GeV}} \right)^2 {\rm
 Gauss}\,.
\label{GUTbound}
\end{equation} 
Although the approximations underlying the semiclassical calculation
break down before such fields are reached, this is probably an
overestimate of the maximum possible field.  Arguments similar to
those leading to Eq.~(\ref{electroweakfield}) indicate that there
should be a local restoration of the GUT symmetry when $B \sim
\alpha^{3/2} M_{\rm mon}^2$.  (The resulting regions of symmetric
vacuum can be viewed as condensed monopole and antimonopole cores.)
With the unbroken symmetry enlarged to a simple non-Abelian group,
magnetic flux is not conserved and a coherent long-range magnetic
field can no longer be sustained.

There are several approaches one could adopt for studying QCD in a
strong magnetic field.  At the hadronic level, one can study the
effect of a magnetic field on hadron spectra \cite{Proton} and the
nuclear equation of state \cite{HadronicMatter}, based on the large
anomalous magnetic moments of hadrons.  Alternatively, one can take a
diagrammatic approach, and compute the quark condensate in a large
magnetic field by resumming diagrams \cite{ShushpanovSmilga}.  The
effect of a magnetic field has also been investigated in the
Nambu-Jona-Lasinio model \cite{KlevanskyLemmer, Schramm, Ebert} and in
an instanton-inspired model for chiral symmetry breaking
\cite{Schramm}.  In contrast, we study the problem using the quark
model.  The advantages of this approach are simplicity and a clear
physical picture of the QCD vacuum.

Throughout this paper we set $\hbar = c = 1$.  The
conversion factor is $1 \,\, {\rm GeV}^2 / (\hbar c)^{3/2} = 1.44
\times 10^{19} \,\, {\rm Gauss}$.  In the introduction we have used
Gaussian units, but in the remainder of this paper we will exclusively
use Heaviside-Lorentz units: $B_{\rm Gaussian} = \sqrt{4 \pi} B_{\rm
Heaviside-Lorentz}$ and $q_{\rm Gaussian} = q_{\rm Heaviside-Lorentz}
/ \sqrt{4\pi}$.  Thus, for example, in the remainder of this paper the
charge of an up quark is $q = {2 \over 3} \cdot \sqrt{4 \pi \alpha}$.

An outline of this paper is as follows.  In section 2 we study the
behavior of a $q\bar{q}$ pair in a strong magnetic field, with the
help of a potential model for the quark-antiquark interaction.  In
section 3 we estimate the strength of the $q\bar{q}$ condensate in
magnetic fields somewhat above the QCD scale.  In section 4 we study
the condensate in the regime of large fields, where perturbative QCD
is applicable.  Section 5 contains a summary and some concluding
comments.  For completeness, in the appendix we compute the response
of the QCD vacuum to weak magnetic fields by performing a pion loop
calculation.

\section{Mesons in a strong magnetic field}

In a strong magnetic field quarks follow Landau orbits in the
directions transverse to the magnetic field.  These have a
characteristic radius $R = 1 / \sqrt{qB}$, so that for $B \gtrsim 1
\,\, {\rm GeV}^2$ the quarks can be localized in the two transverse
directions to distances shorter than the QCD scale.

Moreover, there is no energy cost associated with this localization.
Intuitively, this is because the quark kinetic energy is canceled when
the magnetic moment of the quark lines up with the magnetic field.
More precisely, the energy levels for a Dirac particle in a background
magnetic field are
\be
E(n,\sigma,p_z) = \pm \sqrt{\vert qB \vert (2n + \sigma + 1) + p_z^2 + m^2}\,.
\ee
Here $n = 0,1,2,\ldots$ labels the Landau levels, $\sigma = \pm 1$
specifies the spin orientation, and $p_z$ is the momentum in the $z$
direction.  Thus in the lowest Landau level, with an appropriate spin
orientation, the quark behaves just like a relativistic particle in
$1+1$ dimensions.

One might expect that this localization enhances the attraction
between a color-singlet quark and antiquark to the point where the
energy of a $q\bar{q}$ state becomes negative.  This would
signal an instability with respect to formation of a spin-polarized
$q\bar{q}$ condensate.

To address this issue, we wish to estimate the energy of a $q\bar{q}$
state in a strong magnetic field.  We do this by adopting a potential
model for the $q\bar{q}$ interaction \cite{PotentialReview}.  That
is, we will take the Hamiltonian for a $q\bar{q}$ state to be given by
the quasi-relativistic expression
\be
\label{3dHam}
H = 2 \sqrt{{\bf p}^2 + m^2} + V(r)\,.
\ee
A wide variety of potentials have been discussed in the literature; we will
use the Cornell potential \cite{Cornell}
\be
\label{3dPot}
V(r) = A r - {\kappa \over r} + C  \,.
\ee
We will be focusing on $u\bar{u}$ or $d\bar{d}$ states, and
so use parameters \cite{Fulcher}
\bea
\nonumber
A & = & 0.203 \,\, {\rm GeV}^2 \\
\label{parameters}
\kappa & = & 0.437 \\
\nonumber
m & = & 0.150 \,\, {\rm GeV} \\
\nonumber
C & = & -0.599 \,\, {\rm GeV}
\eea
chosen to fit the the spectrum of light mesons.

In a strong magnetic field this three-dimensional model should reduce
to an effective one-dimensional problem.  However, at distances
shorter than the magnetic length $R$ the problem again becomes
three-dimensional.  We can take this into account by cutting off our
one-dimensional potential at short distances.  Thus we study the
one-dimensional problem
\bea
\label{1dHam}
\vphantom{\Biggl(}\displaystyle
H & = & 2 \sqrt{p_z^2 + m^2} + V(z) \\
\noalign{\vskip 2mm}
\nonumber
V(z) & = & \left\lbrace
\begin{array}{ll}
\vphantom{\Biggl(}\displaystyle
AR - {\kappa \over R} + C & \qquad \vert z \vert < R \\[1mm]
\noalign{\vskip 1mm}
\vphantom{\Biggl(}\displaystyle
Az - {\kappa \over z} + C & \qquad \vert z \vert > R \, .
\end{array}
\right.
\eea
By considering a Gaussian trial wavefunction, one can easily see that
as $R \rightarrow 0$ the spectrum of this Hamiltonian is unbounded
from below.

To estimate the energy levels of the Hamiltonian (\ref{1dHam}) we use
a WKB approximation \cite{Brau}.  The classical turning points are at
$z = \pm L$, where $V(L) = E - 2 m$.  The WKB quantization condition
$\oint p_z dz = 2 \pi (n + {1 \over 2})$ becomes
\be
\label{WKB}
\int_0^L dz \sqrt{(V(z) - E)^2 - 4 m^2} = \pi \left(n + {1 \over 2}\right)
\qquad n = 0,1,2,\ldots
\ee
The resulting ground state energy is shown in Fig.~1.  Note that the
energy is negative for $qB \gtrsim 2 \,\, {\rm GeV}^2$.  We expect the WKB
approximation to give a reasonable estimate for the ground state
energy in this regime.

\begin{figure}
\epsfig{file=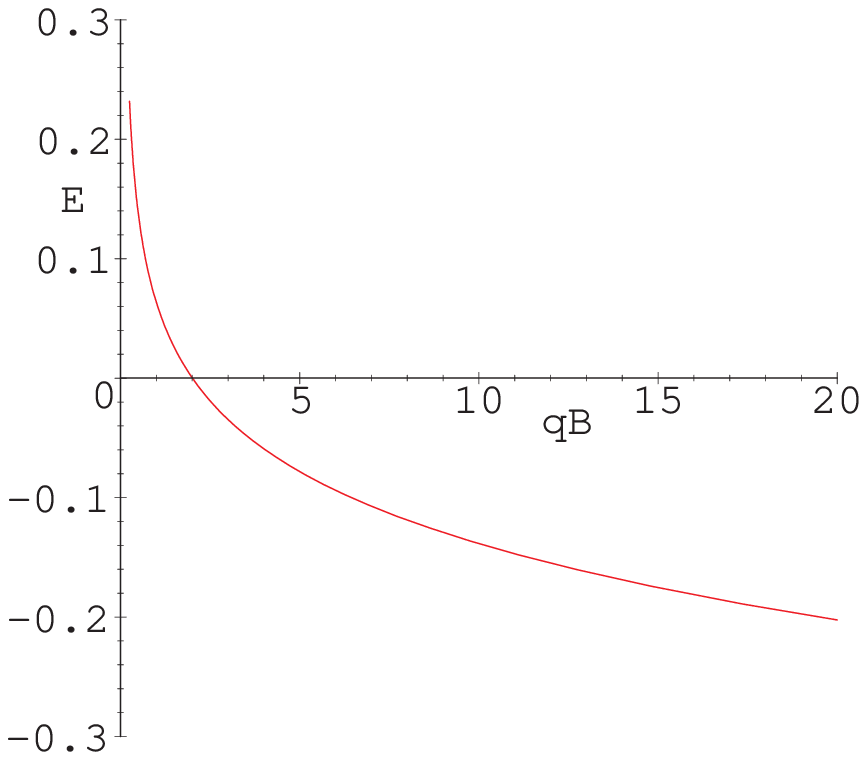}
\caption{Ground state energy as a function of $qB$.  $E$ is in units of GeV,
$qB$ is in units of GeV${}^{2}$.}
\end{figure}

\begin{figure}
\epsfig{file=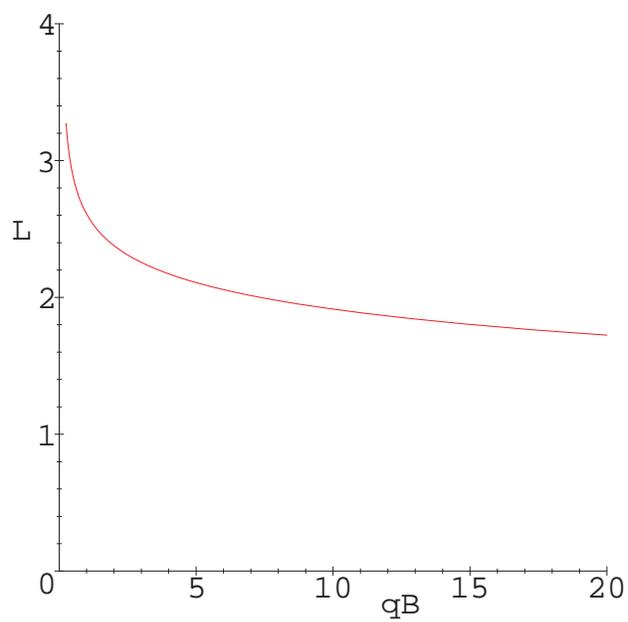}
\caption{Turning point $L$ as a function of $qB$.  $L$ is in units of GeV${}^{-1}$,
$qB$ is in units of GeV${}^{2}$.}
\end{figure}

\begin{figure}
\epsfig{file=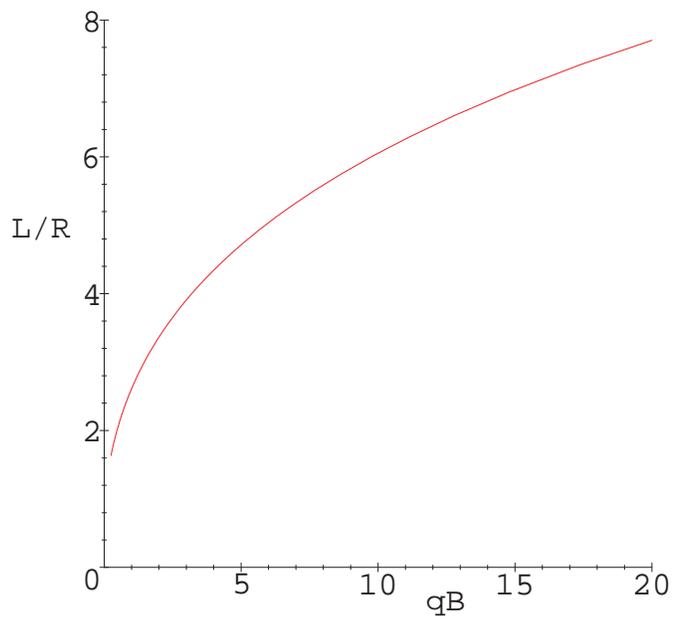}
\caption{$L/R$ as a function of $qB$ (measured in units of GeV${}^{2}$).}
\end{figure}

The semiclassical turning point $L$ is shown in Fig.~2.  Our reduction
to one dimension only makes sense if $L$ is large compared to $R$.  As
can be seen in Fig.~3 this condition is reasonably well satisfied in
the regime of interest.  Note that $L$ decreases as the magnetic field
gets bigger.  This means that at sufficiently large magnetic fields
the $q\bar{q}$ bound state is driven into a short-distance regime
where perturbative QCD can be applied.  This regime is discussed in
more detail in section 4.

\section{Meson condensation}

When the magnetic field is sufficiently large, $q\bar{q}$ pairs will
start to condense.  We first consider a single flavor, and discuss
condensation of $u\bar{u}$ pairs.  Condensation occurs when $qB
\gtrsim 2 \,\, {\rm GeV}^2$, which for a $u\bar{u}$ composite means
\be
\label{CriticalField}
B > B_{\rm crit} \approx {2\,\,{\rm GeV}^2 \over {2 \over 3} 
   \cdot \sqrt{4 \pi \alpha}} \approx 10 \,\, {\rm GeV}^2\,.
\ee
The quark magnetic moments line up with the magnetic field, so $B$
is increased by the formation of the condensate.  This would
seem to make the vacuum unstable, but eventually the $u\bar{u}$ pairs
will start to interact, and this effect presumably stabilizes the
system.

Because the $u\bar{u}$ pairs are color singlets, they will not
interact strongly until their wavefunctions begin to overlap.  Hence, 
for $B > B_{\rm crit}$ pair production should proceed unimpeded until
the density of $u\bar{u}$ pairs reaches a value of roughly
\be
\label{NumberDensity}
\rho = {1 \over \pi R^2 L} = {q B \over \pi L}\,.  
\ee
Once this density is attained, QCD interactions between pairs will
tend to suppress further growth in the condensate.  We will use
Eq.~(\ref{NumberDensity}) as our estimate for $\rho$, although the
actual value that emerges from the interplay of magnetic and QCD
effects will presumably have a somewhat more complicated dependence on
$B$.

Treating the quark and antiquark as elementary Dirac fermions, the
magnetic moment of a pair is $\mu = q / m$ and the magnetization is
\be
\label{magnetization}
M = \mu \rho = {q^2 B \over m \pi L}\,.
\ee
To evaluate this, note that Fig.~2 shows that for $B \gtrsim B_{\rm
crit}$ the length $L$ is slowly varying, with $L \approx 2 \,\, {\rm
GeV}^{-1}$.  The question of what value to use for the mass is a bit
more subtle.  At zero magnetic field one uses constituent quark masses $m
\approx 300 \, {\rm MeV}$ to estimate magnetic moments, although for
extremely large magnetic fields, where $R$ is very small, a current
quark mass may be more appropriate.  Using the $u$-quark charge in
Eq.~(\ref{magnetization}) gives
\be
\label{StrongField}
M = 0.022 \left({300\,{\rm MeV} \over m}\right) 
\left({2\,\,{\rm GeV}^{-1} \over L}\right) B\,.
\ee
Hence, we expect that in the regime $B \gtrsim B_{\rm crit}$ the
magnetization will be small compared to $B$, so that we are justified
in ignoring the back-reaction of the magnetization on the strength of
the condensate.

We now consider the effects of the other quark flavors.  If there were
no interaction between quarks of different flavors, extension of the
above analysis to $d$ quarks would predict that a $d \bar d$
condensate forms at a critical field which is twice as large as for
$u\bar u$ pairs, and with a magnetization that is one quarter as large
for a given $B$.  However, the different condensates will
interact with each other, so that an increase in the condensate of one
flavor will tend to cause a compensating decrease in the other
condensates.  Because of the relatively large $u$ quark electric
charge, a $u \bar u$ condensate is energetically favored over $d \bar
d$ or $s \bar s$, while the heavier quarks are suppressed by their
mass.  Hence we expect the condensate to be dominated by $u\bar{u}$
pairs.

\section{Condensation in the perturbative regime}

When the magnetic field is far above the QCD scale the $q\bar{q}$
composite is driven into a short-distance regime where perturbative
QCD can be applied.  In this section we discuss the magnetization in
this perturbative regime.

At short distances we should replace the Cornell potential of
Eq.~(\ref{3dPot}) with the potential from one-gluon exchange, 
\bea
\label{OneGluonPotential}
&& V(r) =  - {4 \over 3} \, {\alpha_s(r) \over r} 
  \equiv - {A \over r \log (1/\Lambda r)} \\
\nonumber
&& A = {8 \pi \over 3 b_0} = {8 \pi \over 11 N_c - 2 N_f}\,.
\eea
The WKB quantization condition for this potential is still given by
Eq.~(\ref{WKB}).

If one makes the approximation of neglecting the quark mass, the WKB
integral can be evaluated analytically to obtain a relation between
the turning point $L$ and the radius $R = 1 / \sqrt{q B}$.
\be
\label{TurningPoint}
{A \over \log(1/L\Lambda)} + A \log \log {1 \over L \Lambda} =
{A \over \log(1/R\Lambda)} + A \log \log {1 \over R \Lambda} - {\pi \over 2}
\,.
\ee
For $R \rightarrow 0$ this, together with Eq.~(\ref{magnetization}),
gives a magnetization
\be
\label{LargeField}
M = {q^2 B \over m \pi L} \approx {q^2 B \Lambda \over m \pi}
\left({q B \over \Lambda^2}\right)^{{1 \over 2} \exp(-\pi/2A)} \,.
\ee
To evaluate the exponent in the last factor we set $N_c=3$ and take
$N_f$ to be the number of quark flavors that are lighter than the mass
scale set by $B$.  The dependence on $N_f$ is actually rather weak,
with any value between 2 and 6 yielding an exponent of about 0.1.

Of course these calculations are only valid if the turning point $L$
is small enough to trust the potential of Eq.~(\ref{OneGluonPotential}).
This requires magnetic fields that are far larger than those we have
considered so far.  For example, $L < {1 \over 3} \, \Lambda^{-1}$ for a
$u\bar{u}$ composite requires $B > 5 \times 10^{12} \, \Lambda^2$;
although several orders of magnitude above the electroweak scale, this
is still far below the upper limit of Eq.~(\ref{GUTbound}).

We expect that our approximations give the right qualitative behavior
of the magnetization at strong fields.  However an accurate
quantitative calculation calls for more sophisticated techniques than
we have employed here.  For one thing, the WKB estimate of
Eq.~(\ref{TurningPoint}) gets worse as the magnetic field increases,
since the condition for the validity of WKB, $|\partial
p_z^{-1}/\partial z| \ll 1$, is violated near $z = R$.  (The left hand
side grows logarithmically as $R \rightarrow 0$.)  A more serious
concern is that as $R \rightarrow 0$ we should really treat the
$q\bar{q}$ composite in a fully relativistic manner, e.g., by solving
a Bethe-Salpeter equation.

Furthermore, at sufficiently high fields the nonlinearities become
important enough that we must take into account the back-reaction of
the magnetization on the condensate.  These nonlinearities can arise
from several sources.  First, there is the explicit nonlinearity in
Eq.~(\ref{LargeField}), which shows that $M/B$ includes a factor
that grows as a small power of the magnetic field.  Next, we expect
the effective mass of the quarks to decrease as $B$ grows, also
leading to an increase in $M/B$.  A third possible source is the
corrections to our estimate Eq.~(\ref{NumberDensity}) for the density
of condensed pairs; these should also give an increase in
$M/B$ at stronger fields.

\section{Conclusions}

In this paper we have used a quark model approach to study the
behavior of QCD in the presence of a strong magnetic field.  In the
presence of such a field the quarks can be localized in the two
transverse directions with no cost in energy.  This enhances the
quark-antiquark attraction to such an extent that the binding energy
can compensate for the mass, thus making $u\bar u$ pair production
energetically favorable, if $B$ is greater than a critical value of
about 10 GeV$^2$.

In the language of field theory, this pair production corresponds to
the formation of a chiral symmetry breaking $u\bar u$ condensate.  Of
course, even in the absence of a magnetic field, nonperturbative QCD
dynamics produce a nonzero quark condensate that breaks chiral
symmetry.  However, the zero-field and the high field condensates
differ in some significant aspects.

At zero field the condensate is Lorentz invariant.  In particular,
$\langle \bar q \sigma^{\mu\nu} q \rangle = 0$.  By contrast, the
quark pairs produced by a critical magnetic field are polarized along
the direction of the magnetic field.  For a field directed along the
$z$-direction, this corresponds to a condensate with $\langle \bar qq
\rangle \approx \langle \bar q \sigma^{12} q \rangle$.\footnote{The
possibility of a $\langle \bar q \sigma^{\mu\nu} q \rangle$ condensate
was raised in \cite{Schramm}.}

The flavor properties of the two condensates are also quite different.
The zero temperature, zero field condensate is, to a good
approximation, flavor SU(3) symmetric, with $ \langle \bar u u \rangle
\approx \langle \bar d d \rangle \approx \langle \bar s s \rangle$.
This is not the case in the presence of a super-critical magnetic
field, since the production of $u \bar u$ pairs is energetically
favored over that of $d \bar d$ and $s \bar s$ pairs.

Finally, the zero field and high field condensates differ in
magnitude.  The former is of the order of $\Lambda_{\rm QCD}^3 \sim
(.25 \, {\rm GeV})^3$.  This should be compared with our estimate,
Eq.~(\ref{NumberDensity}), for the density of quark pairs.  This
density increases faster than linearly with $B$, but even at the
critical field for $u\bar u$ production we have $\rho \sim (0.7 \, {\rm
GeV})^3$.

We would like to understand the transition between the zero field and
high field regimes.  As $B$ is increased from zero, its initial effect
is to gradually polarize the QCD chiral condensate.  For weak fields,
the relevant degrees of freedom are the Goldstone modes of the
condensate.  These can be studied by using a low-energy chiral
effective Lagrangian.  This leads to a pion-loop calculation, which we
review in the appendix, that gives a magnetization
\be
\label{WeakField2}
M \sim
\left\lbrace
\begin{array}{ll}
\vphantom{\Biggl(}\displaystyle
{7 e^4 B^3 \over 1440 \pi^2 m_\pi^4} 
  & \quad \hbox{\rm for $\vert eB \vert \ll m_\pi^2$} \\ 
\noalign{\vskip 1mm}
\vphantom{\Biggl(}\displaystyle
{e^2 B \over 48 \pi^2} 
     \log {e B \over m_\pi^2} & \quad \hbox{\rm for
$\vert eB \vert \gg m_\pi^2$} \, .
\end{array}
\right.
\ee
Using a similar approach and working in the chiral limit, Shushpanov
and Smilga \cite{ShushpanovSmilga} find that the overall magnitude of
the quark condensate is enhanced by a factor
\begin{equation}
{\Sigma(B) \over \Sigma(0)} = 1 + {eB \ln 2 \over 16 \pi^2 F_\pi^2} + \dots \, .
\end{equation}

As the field increases, higher order terms in the chiral Lagrangian
become important.  In any case, the chiral Lagrangian must be
abandoned in favor of a description in terms of quarks for $eB \gtrsim
(4 \pi F_\pi)^2 \approx 1 \,\, {\rm GeV}^2$.  It would be desirable to
understand this transition region between the weak and strong field
regimes.

Within the strong field regime, our quark model calculations give an
estimate for the magnetization, given in Eqs.~(\ref{StrongField}) and
(\ref{LargeField}).  One would like a clearer understanding of how the
interplay of electromagnetic and QCD effects determines the density of
quark pairs, thus leading to an improvement upon these estimates.  The
development of improved techniques for performing precise calculations
in this regime is clearly needed.  Finally, the relation between our
methods and the more field theoretic approaches to strong field chiral
symmetry breaking of \cite{ShushpanovSmilga} should be better
understood.

\bigskip
\goodbreak
\centerline{\bf Acknowledgements}
We are grateful to Al Mueller, D.~T.~Son and Mal Ruderman for valuable
discussions.  This work is supported in part by the US DOE under contract
DE-FG02-92ER40699 and in part by KOSEF 1998 Interdisciplinary Research
Grant 98-07-02-07-02-5 (K.L.).

\appendix
\section{Weak-field results}

The response of the QCD vacuum to a weak magnetic field can be
computed by using a chiral effective Lagrangian
\cite{ShushpanovSmilga}.  One can integrate out the matter fields to
obtain an effective electromagnetic action $S_{\rm eff}({\bf B})$.  At
leading order the matter contribution to this effective action,
$S_{\rm eff}^{\rm matter}({\bf B})$, arises from a single pion loop.
Thus, we consider a complex scalar field coupled to electromagnetism
with action
\begin{equation}
    S = \int d^4x \left[- {1 \over 4 } F_{\mu\nu} F^{\mu\nu} - \vert
    (\partial_\mu - i eA_\mu) \phi \vert^2 - m^2 \vert \phi \vert^2
    \right] \, .
\end{equation}
Schwinger's classic calculation \cite{Schwinger}
for a uniform magnetic field then gives
\begin{eqnarray}
  S_{\rm eff}^{\rm matter} & = & i \log \det \left(-{\cal D}_\mu {\cal
   D}^\mu + m^2\right) \\ 
   & = & -i \int_{\epsilon^2}^\infty {ds \over s} \, {\rm Tr} \, e^{-i s
   \left(-{\cal D}_\mu {\cal D}^\mu + m^2\right)} \\ 
    & = & {1 \over 16 \pi^2} \int d^4x \int_{\epsilon^2}^\infty {ds \over
    s^3} \, e^{-sm^2} {esB \over \sinh esB}  \, .
\end{eqnarray}
Expanding this in powers of $B$, one finds both a quartic and a
logarithmic divergence.  The former is a contribution to the vacuum
energy, while the latter is the one pion loop contribution to the
renormalization of the electric charge.  Subtracting these divergences
gives the renormalized matter effective action
\begin{equation}
\label{RenormalizedAction}
  S_{\rm eff}^{\rm matter} = {1 \over 16 \pi^2} \int d^4x \int_0^\infty
  {ds \over s^3} \, e^{-sm^2} 
  \left({esB \over \sinh esB} - 1 + {1 \over 6} e^2 s^2 B^2 \right)
\end{equation} 
where $e$ is the renormalized electric charge.  

The magnetization ${\bf M} = {\bf B} - {\bf H}$ is given by 
\begin{equation}
  {\bf M} = {\bf B} + {\delta S_{\rm eff} \over \delta {\bf B}} =
  {\delta S_{\rm eff}^{\rm matter} \over \delta {\bf B}}\,. 
\end{equation}
Substituting the result from Eq.~(\ref{RenormalizedAction}) then leads
to 
\be
\label{WeakField}
   M = {\partial {\cal L}_{\rm eff}^{\rm matter} \over \partial B} \sim
   \left\lbrace
   \begin{array}{ll}
\vphantom{\Biggl(}\displaystyle
   {e^2 B \over 16 \pi^2} \left({7 e^2 B^2 \over 90 m^4} + {\cal
    O}\bigl((eB/m^2)^4\bigr)\right) & \quad \hbox{\rm for $\vert eB \vert
   \ll m^2$} \\ 
   \noalign{\vskip 1mm}
\vphantom{\Biggl(}\displaystyle
   {e^2 B \over 48 \pi^2} \log {e B \over m^2} & \quad \hbox{\rm for
   $\vert eB \vert \gg m^2$} \, .
\end{array} \right.   
\ee
Since $M > 0$, the QCD vacuum is paramagnetic.  

The effects of higher order terms in the chiral Lagrangian have been
studied \cite{AgasianShushpanov}.  These become large for 
$eB \sim (4 \pi F_\pi)^2 \approx 1 \,\, {\rm GeV}^2$, at which point
the chiral Lagrangian approximation breaks down. For such 
fields the effective photon coupling 
\bea 
   {1 \over e_{\rm eff}^2} & =
    & {1 \over e^2} \left[ 1 - {\partial^2 {\cal L}_{\rm eff}^{\rm matter}
    \over \partial B^2} \right]\\ & \approx & {1 \over e^2} - {1 \over 48
    \pi^2} \log {e B \over m^2} \qquad \hbox{\rm for $\vert eB \vert \gg
   m^2$}  
\eea 
is still small, thus justifying our neglect of the quantum
fluctuations in the electromagnetic field.


\end{document}